\documentstyle[twoside, epsfig]{article}

\input ibvs2.sty

\begin{document}

\IBVShead{5775}{29 May 2007}

\IBVStitle{The ultra-compact binary candidate}
\IBVStitle{KUV\,23182+1007 is a bright quasar}

\IBVSauth{Southworth, J.$^1$; Schwope, A.$^2$; G\"ansicke, B. T.;$^1$ Schreiber, M.\ R.$^3$}

\IBVSinst{Department of Physics, University of Warwick, Coventry, CV4 7AL, UK, email: j.k.taylor@warwick.ac.uk, Boris.Gaensicke@warwick.ac.uk}
\IBVSinst{Astrophysikalisches Institut Potsdam, An der Sternwarte 16, 14482 Potsdam, Germany}
\IBVSinst{Departamento de Fisica y Astronomia, Universidad de Valparaiso, Avenida Gran Bretana 1111, Valparaiso, Chile}

\SIMBADobjAlias{KUV 23182+1007}{[DWS97] Peg 5}
\SIMBADobjAlias{KUV 23061+1229}{WD 2306+122}
\GCVSobj{ES Cet}
\IBVStyp{CWA}
\IBVSkey{spectroscopy}

\IBVSedata{23182spectrum.dat}
\IBVSdataKey{23061spectrum.dat}{KUV 23061+1229}{spectrum}
\IBVSedata{23182spectrum.dat}
\IBVSdataKey{23182spectrum.dat}{KUV 23182+1007}{spectrum}

\IBVSabs{KUV 23182+1007 was identified as a blue object in the Kiso UV Survey in the 1980s. Classification-dispersion}
\IBVSabs{spectroscopy showed a featureless continuum except for a strong emission line in the region of He II 4686 A. This is}
\IBVSabs{a hallmark of the rare AM CVn class of cataclysmic variable star, so we have obtained a high-S/N blue spectrum of this}
\IBVSabs{object to check its classification. Instead, the spectrum shows a strong quasar-like emission line centred on 4662 A.}
\IBVSabs{Comparison with the SDSS quasar template spectra confirms that KUV 23182+1007 is a quasar with a redshift of z = 0.665.}

\begintext

{\it KUV\,23182+1007 was identified as a blue object in the Kiso UV Survey in the 1980s. Classification-dispersion spectroscopy showed a featureless continuum except for a strong emission line in the region of He II 4686\,\AA. This is a hallmark of the rare AM\,CVn class of cataclysmic variable star, so we have obtained a high-S/N blue spectrum of this object to check its classification. Instead, the spectrum shows a strong quasar-like emission line centred on 4662\,\AA. Comparison with the SDSS quasar template spectra confirms that KUV\,23182+1007 is a quasar with a redshift of $z = 0.665$.}

\bigskip

The Kiso Ultraviolet Survey (Noguchi, Maehara \& Kondo 1980; Kondo et al.\ 1984) identified 1186 objects with blue colours in a set of fields observed using the 1.0\,m Schmidt telescope of Kiso Observatory. Classification-dispersion spectroscopy of these objects were presented in a series of papers by Wegner and colleagues. The spectra of three objects, KUV\,01584$-$0939, KUV\,23182+1007 KUV\,23061+1229, were given by Wegner, Boley \& Swanson (1987) and Wegner \& McMahan (1988). All three of these showed an interesting strong emission in the region of the He\,II 4686\,\AA\ spectral line.

\medskip

However, confusion arose between the objects KUV\,23182+1007 and KUV\,23061+1229 in Wegner \& McMahan (1988). In that work, both objects were found to have He\,II 4686\,\AA\ emission lines (with some night-to-night variability noted), but the names in the figure titles and figure captions were in mutual disagreement. Koester et al.\ (2001) have since found that KUV\,23061+1229 is a white dwarf of type DA.

\medskip

Strong He\,II emission is a characteristic of the rare AM\,CVn class of cataclysmic variable star (Warner 1995; Southworth et al., 2006). These objects are particularly interesting ultra-short period helium-rich systems which are thought to be interacting binaries composed of two degenerate objects, the mass donor being a helium white dwarf. KUV\,01584$-$0939 has since been confirmed to be an AM\,CVn star (Warner \& Woudt 2002; Espaillat et al.\ 2005), and is included in the {\em General Catalogue of Variable Stars} under the name ES\,Ceti.

\medskip

As very few AM\,CVn systems are known we have obtained a spectrum of the second of the objects, KUV\,23182+1007, in order to investigate its classification as a cataclysmic variable. We also obtained a spectrum of KUV\,23061+1229 in order to confirm that it is a white dwarf and to fully clear up the confusion over the identities of these two objects. For these observations we adopted the object identifications and sky co-ordinates as given by the CDS {\it Simbad} tool\footnote{\tt http://simbad.u-strasbg.fr/simbad/sim-fid}.

\medskip

\IBVSfig{10cm}{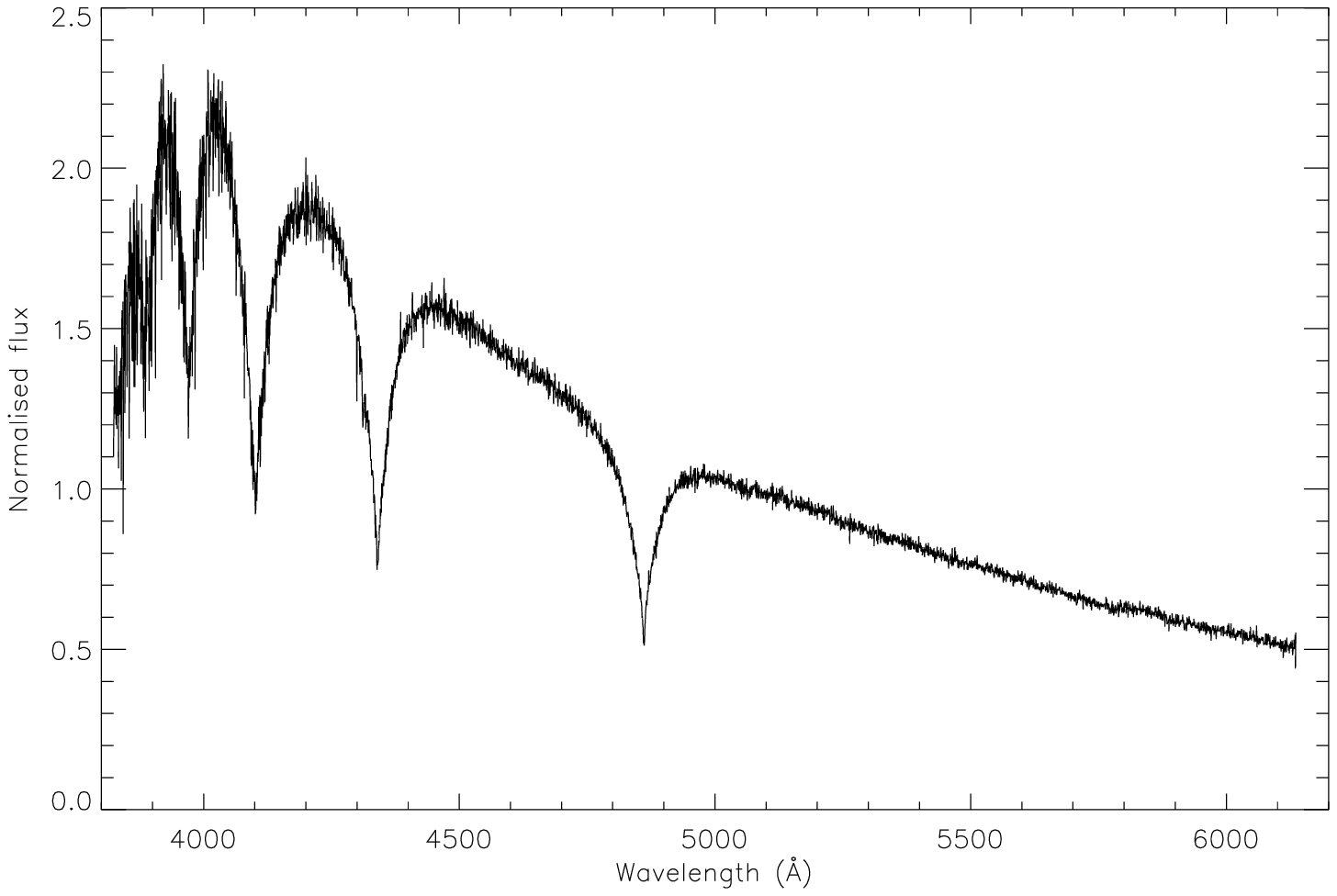}{Magellan/LDSS3 spectrum of the second AM\,CVn candidate, KUV\,23061+1229, confirming that this object is a DA white dwarf.}
\IBVSfigKey{spec23061.eps}{KUV 23061+1229}{spectrum}

Two consecutive long-slit spectra of KUV\,23182+1007, immediately followed by one spectrum of KUV\,23061+1229, were obtained on the night of 2007 May 19. We used the LDSS3 spectrograph attached to the 6.5\,m Magellan Clay telescope at Las Campanas Observatory, Chile. The VPH\_Blue grism was used along with a slit width of 0.75$^{\prime\prime}$, giving a useful wavelength coverage of 4000--6130\,\AA\ (depending on brightness) at a reciprocal dispersion of 0.68\,\AA\,px$^{-1}$. From the arc lamp and sky lines we estimate a resolution of approximately 2\,\AA. Wavelength and flat-field calibration was achieved using observations of helium/neon/argon and quartz lamps, taken immediately after the science spectra and at the same sky position. The two science spectra of KUV\,23182+1007 have been combined and rebinned to increase the signal-to-noise ratio, resulting in a single spectrum with a reciprocal dispersion of 2\,\AA\,px$^{-1}$. The effective midpoint of this observation is HJD 2\,454\,240.88628. The midpoint of the spectrum of KUV\,23061+1229 occurred at HJD 2\,454\,240.90236.

\medskip

The spectrum of KUV\,23061+1229 (Fig.\,1) is clearly that of a DA white dwarf, in agreement with the results of Koester et al.\ (2001) and its inclusion in the white dwarf catalogue of McCook \& Sion (1999). We have therefore adopted the atmospheric parameters found by Koester et al.\ (2001) to calculate a model spectrum (G\"ansicke, Beuermann \& de Martino 1995) of KUV\,23061+1229 and used this to divide out the wavelength-dependent response function of the spectrograph from the spectrum of KUV\,23182+1007.

\medskip

The KUV\,23182+1007 spectrum is plotted in Fig.\,2 and shows a strong emission line at 4660\,\AA\ which we identify to be the Mg 2800\,\AA\ line which is a characteristic feature of quasar spectra. In Fig.\,2 we have also plotted a template quasar spectrum\footnote{The spectrum was obtained from {\tt http://www.sdss.org/dr5/algorithms/spectemplates/spDR2-029.fit}} from the {\it Sloan Digital Sky Survey} to which we have applied a redshift of $z = 0.665$. It can be seen that several additional quasar emission lines match the spectrum of KUV\,23182+1007, confirming that this object is a bright quasar with a redshift of $z = 0.665$.

\medskip

The large width of the Mg II line (FWHM $\sim$50\AA $\equiv \sim$5000 km\,s$^{-1}$) indicates that KUV\,23182+1007 is a type\,I AGN. Using $\Lambda_{\rm CDM}$ cosmological parameters, the distance modulus is 43.0. With the observed $R$-band apparent magnitude $m_R = 17.5$ (a proxy for the rest-frame $B$-band magnitude) the absolute rest-frame $B$-band magnitude becomes $M_B = -25.5$, which confirms that this object is a quasar with a typical absolute brightness (Veron-Cetty \& Veron 2006).

\medskip

\IBVSfig{10cm}{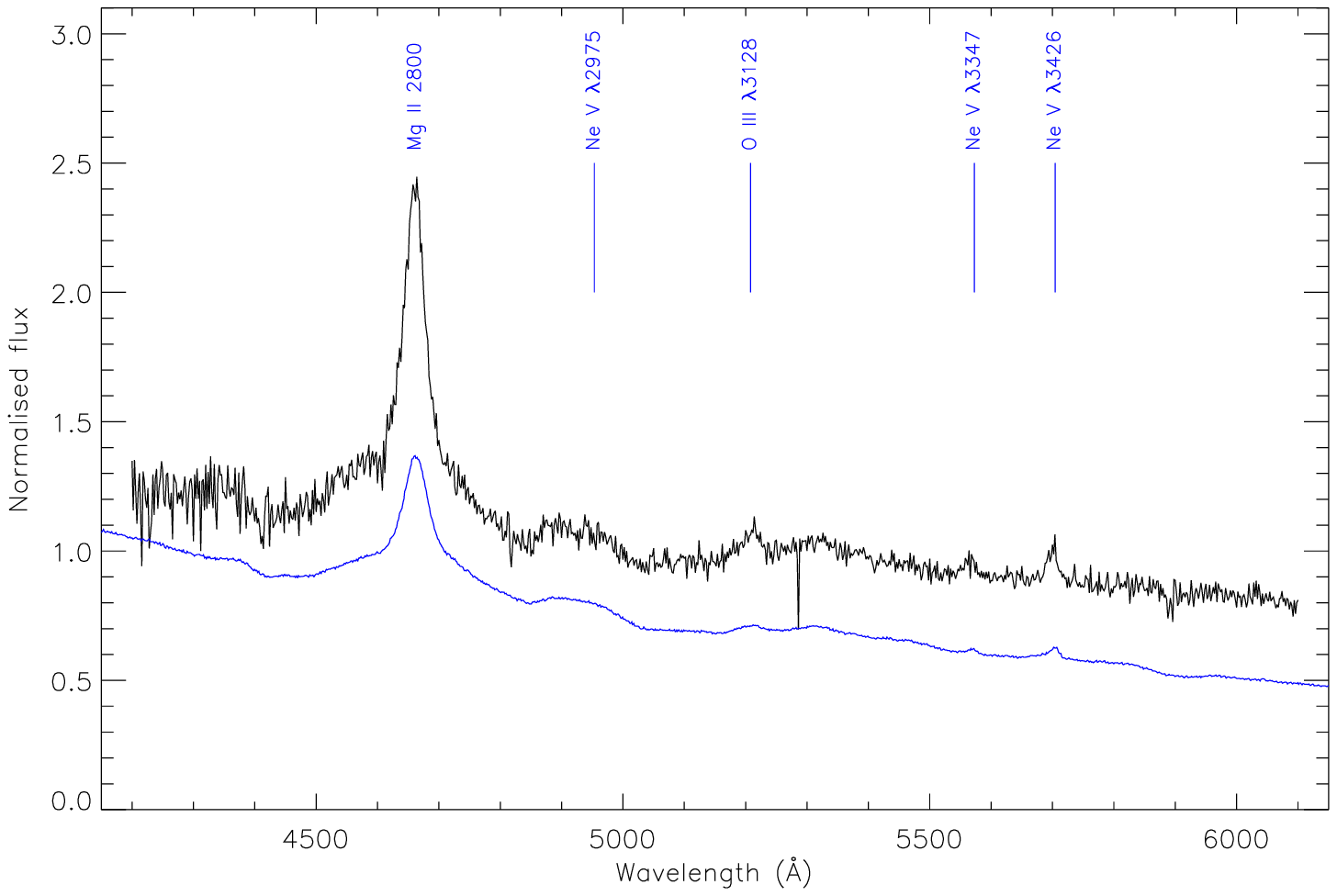}{Magellan/LDSS3 spectrum of the main AM\,CVn candidate, KUV\,23182+1007 (upper solid line), after combining and rebinning. A template quasar spectrum from the SDSS is also shown (lower solid line) after applying a redshift of $z = 0.665$ to the wavelength scale. The stronger quasar emission lines are labelled with their rest wavelengths, taken from Vanden Berk et al.\ (2001).}
\IBVSfigKey{spec23182.eps}{KUV 23182+1007}{spectrum}

As active galactic nuclei are often X-ray sources we have investigated the XMM-Newton and ROSAT databases for sources at the position of KUV\,23182+1007. This region of sky has not been observed using pointed observations by these satellites. However, the ROSAT All-Sky Survey\footnote{The ROSAT All-Sky Survey catalogue can be accessed using the CDS {\it VizieR} service at {\tt http://cdsweb.u-strasbg.fr/viz-bin/VizieR-2?-source=IX/29}} (Voges et al., 1999, 2000) includes an exposure of 444\,s of this position, in which a source RXS\,J232044.6+102354 is detected with a count rate of $0.0249 \pm 0.0094$\,counts\,s$^{-1}$. This is within 6$^{\prime\prime}$ of the position of KUV\,23182+1007, and over 35$^{\prime}$ from the next nearest X-ray source. Given the quoted ROSAT positional error of 15$^{\prime\prime}$, this is a strong detection. The detected X-ray emission is consistent with our identification of KUV\,23182+1007 as a quasar.

\medskip

We have therefore clearly identified that KUV\,23182+1007 is an X-ray emitting quasar with a redshift of $z = 0.665$, and confirmed that KUV\,23061+1229 is a normal DA white dwarf. The classification of KUV\,23182+1007 in {\it Simbad} and catalogues of cataclysmic variables (Downes et al.\ 2001; Ritter \& Kolb 2003) should be corrected. This report is intended to avoid other researchers using valuable telescope time to investigate the basic properties of KUV\,23182+1007.

\references

Downes, R.\ A., Webbink, R.\ F., Shara, M.\ M., Ritter, H., Kolb, U., Duerbeck, H.\ W., 2001, {\it PASP}, {\bf 113}, 764

Espaillat, C., Patterson, J., Warner, B., Woudt, P., 2005, {\it PASP}, {\bf 117}, 189

G\"ansicke, B.\ T., Beuermann, K., de Martino, D., 1995, {\it A\&A}, {\bf 303}, 127

Koester, D., et al., 2001, {\it A\&A}, {\bf 378}, 556

Kondo, M., Noguchi, T., Maehara, H., 1984, {\it Ann.\ Tokyo Astron.\ Obs.}, {\bf 20}, 130

McCook, G.\ P., Sion, E.\ M., 1999, {\it ApJS}, {\bf 121}, 1

Noguchi, T., Maehara, H., Kondo, M., 1980, {\it Ann.\ Tokyo Astron.\ Obs.}, {\bf 18}, 55

Ritter, H., Kolb, U., 2003, {\it A\&A}, {\bf 404}, 301

Southworth, J., et al., 2006, {\it MNRAS}, {\bf 373}, 687

Vanden Berk, D.\ E., et al., 2001, {\it AJ}, {\bf 122}, 549

Veron-Cetty, M.\ P., Veron, P., 2007, {\it A\&A}, {\bf 455}, 773

Voges, W., et al., 1999, {\it A\&A}, {\em 349}, 389

Voges, W., et al., 2000, {\it IAU Circ.}, {\bf 7432}

Warner, B., 1995, {\it Cataclysmic Variable Stars}, Cambridge University Press

Wegner, G., Boley, F.\ I., Swanson, S. R., McMahan, R.\ K., 1987, in IAU Coll.\ 95: Second Conference on Faint Blue Stars, eds.\ A.\ G.\ D.\ Philip, D.\ S.\ Hayes \& J.\ W.\ Liebertm L.\ Davis Press Inc., p.\,501

Wegner, G., McMahan, R.\ K., 1988 ,{\it AJ}, {\bf 96}, 1933

Woudt, P., Warner, B., 2002, {\it PASP}, {\bf 114}, 129

\IBVSedata{5xxx-t1.txt}
\IBVSedata{5xxx-t2.txt}

\IBVSefigure{5xxx-f1.ps}
\IBVSefigure{5xxx-f2.ps}

\end{document}